\documentstyle[epsf]{mn}
 
\def\tdrv#1#2{{d #1\over d #2}}
\def\tdrv2#1#2{{d^2 #1\over d{#2}^2}}

\def\Fsum {\sum_{n=-\infty}^{\infty}}
\def\AAsum {\sum_{\lv=-\infty}^{\infty}}
\def\ldomega {\bf l \cdot \Omega \rm}
\def\ldw{\bf l \cdot w \rm}
\def\lv {\bf l \rm}

\def\ldw {\bf l \cdot w \rm}

\def\Ebar {\langle\langle \dot E \rangle\rangle}

\def\Myr{{\rm\,Myr}}

\def\Gyr{{\rm\,Gyr}}

\def\kms{{\rm\,km\,s^{-1}}}
\def\kpc{{\rm\,kpc}}

\def\msun{{\rm\,M_\odot}}

\def\pc{{\rm\,pc}}

\def\spose#1{\hbox to 0pt{#1\hss}}
\def\lta{\mathrel{\spose{\lower 3pt\hbox{$\mathchar"218$}}
     \raise 2.0pt\hbox{$\mathchar"13C$}}}
\def\gta{\mathrel{\spose{\lower 3pt\hbox{$\mathchar"218$}}
     \raise 2.0pt\hbox{$\mathchar"13E$}}}

\begin{document}
 
\title{Evolution of the Galactic Globular Cluster System}
 
\author[Chigurupati Murali and Martin D. Weinberg]{Chigurupati Murali
	\thanks{Present address: Canadian Institute for Theoretical
	Astrophysics, McLennan Labs, Toronto, ON M5S 1A1, Canada} and
	Martin D. Weinberg\thanks{Alfred P. Sloan Foundation
	Fellow.}\\ Department of Physics and Astronomy, University of
	Massachusetts, Amherst, MA 01003-4525, USA}

\maketitle

\begin{abstract}

	We study the dynamical evolution of disk and halo globular
clusters in the Milky Way using a series of Fokker-Planck calculations
combined with parametric statistical models.  Our sample of 113
clusters with velocity data is predicted to descend from an initial
population of 250 clusters, implying more than a factor of two
decrease in population size due to evolution.

	Approximately 200 of these clusters are in a halo component
and 50 in a disk component.  The estimated initial halo population
follows a coreless $R^{-3.38}$ density profile in good agreement with
current estimates for the distribution of halo field stars.  The
observed core in the present-day distribution of halo clusters results
from the rapid evaporation of clusters in the inner regions of the
Galaxy.  The initial halo population is also predicted to have a
radially biased orbit distribution in rough agreement with the
observed kinematics of halo field stars.  The isotropy of the
present-day halo cluster distribution results from the evaporation of
clusters on elongated orbits.  Similarly, the initial disk component
has a nearly isotropic initial distribution that becomes more
tangentially biased with time.  However, the inferred initial
characteristics of the disk component do not match the kinematics of
the rapidly rotating thin or thick disk stellar populations.  These
characteristics may be more indicative of the flattened halo component
discussed by Zinn (1993).

	Detailed examination of cluster evolution confirms the
importance of disk heating.  Clusters on low-inclination orbits
experience the strongest disk heating because of optimal matches in
resonant frequencies.  Disk heating on high-inclination orbits is
weaker but still dominates over spheroidal heating.  Evaporation times
depend weakly on initial concentration, density and height of
oscillation above the disk.

\end{abstract}


\section{Introduction}
\label{sec:intro}

	Age estimates for globular clusters indicate that they formed
early in the history of the Milky Way and represent `fossil relics' of
the proto-Galaxy (Larson 1990).  Attempting to uncover this history,
researchers have carefully examined a range of properties of the
present-day cluster system, paying particular attention to the cluster
kinematic distribution (e.g. Zinn 1993), mass distribution
(e.g. Harris \& Pudritz 1994), metallicity distribution (e.g. Zinn
1985) and age distribution (e.g. Chaboyer, Demarque \& Sarajedini
1996).  These investigations have provided evidence for both accreted
and native components in the cluster system (Searle \& Zinn 1978) and
correlations between kinematics and metallicity which may trace the
collapse of the Galaxy (Zinn 1985; Armandroff 1989; Zinn 1993).

	Comparisons with other stellar populations have revealed
subtleties in the process of Galaxy formation and evolution.  For
example, in the inner Galaxy, the globular cluster distribution is
flatter than the distribution of halo field stars although their
profiles match well at larger radii.  In addition, the field star
velocity ellipsoid has a strong a radial bias in comparison to the
approximately isotropic cluster velocity ellipsoid (e.g. Ostriker,
Binney \& Saha 1989).  Conversely, Zinn (1985) and Armandroff (1989)
have presented convincing evidence for a high-metallicity disk cluster
population which has broad similarities in kinematics, spatial
distribution and metallicity with the stellar thick disk.
Understanding the origin of these relationships will improve our
picture of the primordial Milky Way.

	At the same time, theoretical interest in globular cluster
evolution has been motivated by the discovery that two-body relaxation
would drive evolution on a time scale that is much less than the age
of a typical cluster (Ambartsumian 1938; Spitzer 1940; Chandrasekhar
1942).  Subsequent research provided an understanding of the
gravothermal instability (Lynden-Bell \& Wood 1968) and the phenomenon
of core collapse (e.g. Cohn 1980).  One of the basic conclusions of
this work on cluster evolution is that relaxation inevitably leads to
evaporation (e.g. Spitzer 1987).  Additional refinements to the
picture of relaxation-driven evolution have been required to account
for a source of energy which halts core collapse (e.g. Henon 1961; Lee
\& Ostriker 1987) and to include tidal influences which arise on a
cluster's orbit in a parent galaxy (Ostriker, Spitzer \& Chevalier
1972; Chernoff, Kochanek \& Shapiro 1986; Weinberg 1994; Gnedin \&
Ostriker 1996; Murali \& Weinberg 1996, hereafter Paper I).

	Recent work on tidal influences has shown that evaporation is
accelerated by the interaction of a cluster with the tidal field
produced by the halo and disk of the Galaxy (Gnedin \& Ostriker 1996;
Paper I).  These studies show that depletion depends strongly on
cluster mass and orbit in the Galaxy.  This suggests that
understanding the initial characteristics of clusters and their
relationship to other stellar populations in the Galaxy requires a
comprehensive description of evolution since the time of formation.
In a related study which supports this view, Murali \& Weinberg (1996;
hereafter Paper II), have demonstrated the importance of evolution in
shaping the M87 globular cluster population and as a partial cause for
the specific frequency conundrum in fundamental plane ellipticals.

	Motivated by these results in the present work, we investigate
the degree to which evolution has shaped the Milky Way cluster
population.  Our approach is to predict the initial spatial and
kinematic distributions of the globular cluster system using the
Fokker-Planck description of cluster evolution discussed in Paper I in
combination with the parametric statistical framework employed in
Paper II.  The predictions describe the initial population of clusters
which evolve quasi-statically through relaxation and tidal heating and
indicate changes which dynamical evolution has wrought on the system
as a whole.  The results also provide a basis for understanding the
primordial relationship of globular clusters to other stellar
populations.

	We first study the evolutionary behavior of clusters which
inhabit the disk and halo of the Galaxy.  The calculations demonstrate
the importance of disk heating on cluster evolution and quantify
dependences on important internal and external parameters, including
orbit in the Galaxy and cluster concentration.  We also examine the
behavior of internal density profiles and mass spectra in evolving
clusters.

	Having considered the detailed physical behavior, we examine
properties of the full cluster population.  We first characterize
properties of the current cluster population and then predict its
initial conditions using the data set compiled by Gnedin \& Ostriker
(1996) and the three-space velocities derived by Cudworth (1993).  The
inferences are derived from both spherical and two-component
disk+sphere models of the cluster distribution.  While several
analyses have shown the cluster system to be approximately spherically
distributed (Chernoff \& Djorgovski 1989; Thomas 1989), other
investigations show two components: a flattened, rapidly rotating
high-metallicity component associated with the Galactic disk and a
spherically distributed low-metallicity component associated with the
Galactic halo (Zinn 1985; Armandroff 1989).  Further subdivisions may
also exist (Zinn 1993; Zinn 1996).  The choice of models reflects the
gross characteristics of the cluster system and allows us to compare
the candidate distributions.  The results predict significant
differences in the initial and present-day cluster populations,
indicating the role of evolution in shaping the present-day cluster
system.  Moreover, neither model is completely successful in
describing the cluster population probably due to the combined effects
of evolution and obscuration.

	The plan of the paper is as follows.  In
\S\ref{sec:investigation}, we summarize the approach and scenario used
throughout the investigation.  The results are presented in
\S\ref{sec:results} and include description of the physical behavior
as a function of orbit, examination of the internal properties of
evolving clusters and analysis and prediction of the initial
conditions of the observed population.  Finally,
\S\ref{sec:discussion} discusses the implications of the results.  The
appendices provide derivations of the models and a discussion of the
statistical procedure.

\section{Scenario and Investigation}
\label{sec:investigation}

\subsection{Model populations}
\label{mod:pop}

	Our fiducial population consists of clusters which formed in a
single episode approximately $11 \Gyr$ ago, the lower limit on cluster
ages estimated from current models of stellar evolution (Chaboyer
1995).  For older ages, the evolution in the properties of the cluster
system is greater than the estimates derived below.  Initial clusters
are assigned $W_0=5$ King model profiles.  Investigation of
concentration dependence in \S\ref{sec:concentration} shows that
evaporation times vary little with $W_0$.  We assume that each cluster
has a Salpeter IMF $m^{-\beta}$ with $\beta=2.35$ and lower mass limit
$m_l=0.1 \msun$.  For this choice, stellar evolution would dominate
for the first \Gyr, roughly corresponding to the main sequence
lifetime of a $2 \msun$ A-star, which we choose as the upper mass
limit, $m_u$.  Following the phase of strong stellar evolution,
relaxation, external heating and, ultimately, core collapse heating
would begin to drive cluster evolution.  We define our zero-population
at this epoch, approximately $10\Gyr$ in the past.

	In light of observational evidence, we adopt a two-component
model of the cluster population consisting of flattened and spherical
distributions.  We employ the commonly used terminology of `disk' and
`halo' cluster to refer to members of these sub-populations.  A `disk'
cluster is most often `metal-rich' with disk kinematics while a `halo'
cluster is most often `metal-poor' with halo kinematics but which may
have high or low orbital inclination relative to the disk.  A
`classic' halo orbit, however, is one of high inclination from the
disk.

\subsection{Cluster evolution}
\label{sec:cl_ev}

	Following formation and early stellar evolution, cluster
evolution is driven by relaxation, the tidal field and binary heating
of the core.  As described in Papers I \& II, the competition between
relaxation and the tidal field of a galaxian spheroid is particularly
important in determining a cluster's evolutionary time scale and
survival history.  The dominant effects of the spheroid on
tidally-limited clusters were found to be heating on low-eccentricity
orbits and tidal truncation on high-eccentricity orbits.

	In the Milky Way, the disk also contributes significantly to
cluster evolution.  For high inclination orbits, a number of
investigations have shown that the compressional shock imparted to a
cluster during its passage through the disk will generally enhance the
evaporation rate (Spitzer \& Chevalier 1972; Chernoff, Kochanek \&
Shapiro 1986; Weinberg 1994; Gnedin \& Ostriker 1996).  For orbits
confined to the disk, oscillations of the cluster about the midplane
transfer energy through resonant stellar orbits.  Below we examine the
effect of the disk on cluster evolution on both low- and
high-inclination orbits (\S\ref{sec:behavior}).  Appendix
\ref{sec:disk} outlines the derivation of the heating rate for disk
oscillations.

\subsection{Orbits}
\label{sec:orbits}

	For eccentric, high-inclination orbits, the orbital phase of
disk passage varies due to the precession of the argument of
perihelion with respect to the plane of the Galaxy because orbits are
not closed in the logarithmic potential.  To remove dependence on
orbital phase, we assume that disk shocking occurs twice per orbital
period at the average orbital radius of the cluster.  For eccentric
low-inclination orbits, we follow the same procedure and also assume
that the vertical motion is separable from the radial and tangential
motion in all orbits.  This allows us to define approximate
three-integral distribution functions (DFs) in terms of algebraic
constants of the motion (\S\ref{sec:distributions}).  With these DFs,
we can follow the evolution of the phase space distribution of
globular clusters using a series of Fokker-Planck calculations
(\S\ref{sec:calculations})

	Orbits in the spheroid are defined using the quantity
$\kappa=J/J_{max}(E)$, the angular momentum relative to maximum for an
orbit of energy E.  The inclination angle, $i$, is defined with
respect to the disk so that $i=0^{\rm o}$ defines an orbit in the
plane of the disk.  Oscillations through the disk are defined by their
oscillation height in multiples of the disk scale height, $z_0$.

\subsection{Tidal limitation}

	Clusters on orbits highly inclined from the disk are tidally
limited by the Galactic spheroid.  While initial cluster densities may
differ from the mean density required by perigalactic tidal
limitation, subsequent evolution during the first gigayear leads
rapidly to tidal truncation or disruption.  Clusters on orbits
confined to the disk may posess limiting radii, $R_c$, smaller than
that implied by tidal limitation in the spheroid (e.g. note the
discrepancy between the observed and predicted tidal radius of M71:
Drukier, Fahlman \& Richer 1993).  This implies that their density is
higher than that required for tidal limitation at given concentration.
However, \S\ref{sec:density} shows that evaporation times vary little
with density for fixed mass.  Therefore, in studying population
evolution, we assume that low-inclination clusters are also tidally
limited.  The limiting radius $R_c$ is determined by the cluster mass,
orbit and the ratio of cluster mean density to the mean density
required for tidal limitation (see \S\ref{sec:disk_parms} and Paper I
for further details).  When clusters are tidally limited, we refer to
the limiting radius as the {\it tidal radius\rm}, $R_t$.

\subsection{Distribution functions}
\label{sec:distributions}

	We use parametric models to define distributions of cluster
orbits and masses in both disk and spheroid populations.  Mass
distributions, $\nu(M)$, are defined using either a power law (Harris
\& Pudritz 1994), a two-component power law or a Gaussian magnitude
distribution (McLaughlin, Harris \& Hanes 1994).  The two-component
power-law mass spectrum is continuous at the break at mass $M_{cut}$.

	Distributions of cluster orbits are defined using well-known
models (Table \ref{tab:models}) which are generalized to provide
kinematic and spatial distributions for populations in the Galactic
potential (\S\ref{sec:MW_model}).  Disk component DFs are defined as
functions of $E_d$, the orbital energy associated with motion in the
disk plane, $J_z^2$, the square of the z-component of the angular
momentum and $E_z$, the orbital energy associated with motion
perpendicular to the disk plane.  The functional form is obtained
using the Mestel disk with no net rotation (Binney \& Tremaine 1987)
combined with an isothermal vertical distribution.  This provides a
family of constant anisotropy, power-law surface density profiles with
power-law index $-(\eta_d-2q_d)$ of infinite range.  The quantities
$\eta_d\equiv v_c^2/\sigma_d^2$ and $q_d$ respectively define the
velocity dispersion, $\sigma_d$, in terms of the (constant) circular
rotation velocity, $v_c$, and degree of anisotropy of the distribution
in the disk plane.  The parameter $\eta_z\equiv\Phi_0/\sigma_z^2$
defines the vertical scale height of the distribution in terms of the
vertical velocity dispersion, $\sigma_z$, and the central potential of
the disk, $\Phi_0$.  Appendix \ref{sec:mestel_disk} discusses the
model family in further detail.

	Spheroidal component distribution functions are defined as
functions of $E$, the total energy of the orbital motion, and $J^2$,
the square of the total angular momentum.  We use two models.  The
first is obtained by adapting the Mestel DF to the spherical case
which we refer to as the {\it Mestel sphere\rm}.  The second is the
Eddington model (Aguilar, Hut \& Ostriker 1988).  The Mestel sphere
provides a family of constant anisotropy power-law space density
profiles with power-law index $-(\eta-2q)$ of infinite range.  The
quantities $\eta\equiv v_c^2/\sigma^2$ and $q$ respectively define the
velocity dispersion, $\sigma$, in terms of the (constant) circular
rotation velocity, $v_c$, and degree of anisotropy of the distribution
as in the disk case.  The Eddington sphere produces a family of
variable anisotropy, power-law space density profiles with core radius
$R_a$ and power-law index $-\eta$ at large radii.  Radial anisotropy
becomes significant beyond the core radius.  Appendices
\ref{sec:mestel_sphere} and \ref{sec:eddington_sphere} provide
derivations and further discussion of these models.

	Complete orbit and mass distributions are given by joint
distributions $\nu(M)\times f(\vec I)$ where $\vec I$ denotes
constants which define a disk or halo orbit.  We assume that disk and
halo components can have different mass spectra.  In the two-component
model, $F$ denotes the fraction of the population belonging to the
spherical component.  The results of \S\ref{sec:behavior} show that
disk cluster evolution varies little with oscillation height.  Since
the cluster population is not large, we therefore determine the
parameter $\eta_z$ for the observed distribution, and keep it fixed
when estimating the initial conditions in order to minimize the number
of parameters.  For the disk distribution, we impose cutoffs in radius
at $R_d=15\kpc$ (Wainscoat et al. 1992) and height at $Z=7.5\kpc$,
where $R_d$ and $Z$ are the radius in and height above the disk,
respectively (see Appendix \ref{sec:coordinates} for listing of
coordinate notations).

\begin{table*}
\caption{Functional forms of models}
\label{tab:models}
\begin{tabular}{lcl}
\multispan3\hfil Mass Models\hfil\\\\
Name&$\nu(M)\propto$&Parameters\\
\hline
power law&$M^{-\alpha}$&$\alpha$\\\\
two-component&$M^{-\alpha_1} (M\leq M_{cut})$&$\alpha_1$,\\
\hfill power law&$M^{-\alpha_2} (M>M_{cut})$&$\alpha_2,M_{cut}$\\\\
Gaussian&$e^{-(V-V_0)^2/2\sigma_V^2}\cdot dV/dM$&$V_0,\sigma_V$\\
\hline\\\\
\multispan3\hfil Distribution Functions \hfil\\\\
Name&$f(E,J^2,E_z)\propto$&Parameters\\
\hline
Mestel disk&$e^{-E_d/\sigma_d^2}J_z^{2q_d}e^{-E_z/\sigma_z^2}$
	&$0\leq\eta_d<\infty$\\
&&$-{1\over 2}\leq q_d<\infty$\\
&&$0\leq\eta_z<\infty$\\\\
Mestel sphere&$e^{-E/\sigma^2}J^{2q}$&$0\leq\eta<\infty$\\
&&$-1\leq q<\infty$\\\\
Eddington sphere&$e^{-E/\sigma^2}e^{-J^2/2r_a^2\sigma^2}$&$0\leq\eta<\infty$\\
&&$0\leq r_a<\infty$\\\\
\hline
\end{tabular}
\end{table*}

\subsection{Statistical procedure}

	A maximum likelihood method is used to estimate the parameters
in Table \ref{tab:models} from the observed cluster data.  Our models
can include 7 parameters for each cluster (mass, 3 spatial coordinates
and 3 velocity coordinates).  However, not all quantities are
available in most cases.  To incorporate all available data, we use a
likelihood technique for incomplete data sets (Little \& Rubin 1987;
Stuart \& Ord 1991).

	The likelihood is constructed in the usual manner as the joint
probability of all observations given the model.  However, for
clusters with unavailable phase space quantities, the probability is
obtained by integrating the distribution function over all dimensions
of unknown information.  This defines the marginal probability of
observing a particular cluster given the model.  Typically, only the
heliocentric radial velocity is known, so we integrate over the
tangential velocities relative to our vantage point to derive the
marginal probability of observing a cluster at a given position with a
given heliocentric radial velocity.  Appendix \ref{sec:likelihood}
discusses the technique in more detail.

	Two types of fit are used to characterize the data.  Fits
without the evolutionary calculations (c.f. \S\ref{sec:calculations})
are used to derive the observed or {\it present-day \rm} properties of
the cluster sample.  Fits based on the evolutionary calculations are
used to derive the most likely initial conditions which produce
today's distribution.  We first consider the present-day
characteristics of the cluster system to serve as a guide to
interpreting the evolutionary case.  The listed uncertainties are
$1-\sigma$ variances computed under the assumption of normally
distributed errors.

\subsection{Data}

	We use the data compiled in Gnedin \& Ostriker (1996) which
consists of 119 objects.  Comparison with the Harris (1996)
compilation shows no obvious systematic differences.  We examine the
distribution of clusters within $65 \kpc$ having masses in the range
$2.0\times 10^4\msun\leq M\leq 2.75\times 10^6\msun$ for a
mass-to-light ratio of 3.  This removes 6 clusters from the original
sample, leaving a total of 113 clusters.  Using the likelihood
procedure outlined above, we also include the three-space velocities
which now exist for about 20 clusters (Cudworth 1993).  This is
referred to as the {\it augmented data set\rm}.

\subsection{Calculations}
\label{sec:calculations}

	We follow the evolution of individual clusters using the
one-dimensional Fokker-Planck approximation (Cohn 1979).  The
calculations include relaxation, external heating due to the
time-varying tidal field of the disk and spheroid, and a
phenomenological binary heating term (Lee et al. 1991).  Spheroidal
tidal heating is included using the implementation discussed in Paper
I.  Disk tidal heating is implemented as a shock on high-inclination
orbits and as an average heating rate on low-inclination, oscillatory
orbits.  The numerical procedure used on low-inclination orbits is the
same used for orbits in the spheroid.  On high-inclination orbits, we
use an analogous, flux-conserving scheme to compute the total change
in the DF due to the passage through the disk.

	Properties of the evolved cluster population are estimated
from a grid of Fokker-Planck calculations performed over a range of
cluster orbits. For the disk clusters, we use a $4\times4\times5$ grid
in apogalactic radius, mass and $\kappa$ to sample the phase space.
Apogalactica are taken in the range $2\leq R_a\leq 8\kpc$, masses in
the range $10^5 \msun\leq M\leq 5\times 10^6 \msun$ and orbits in the
range $0.3\leq\kappa\leq 1.0$.  For the halo clusters, we use a
$5\times4\times5$ grid with $5\leq R_a\leq 15 \kpc$ and the same range
of mass and $\kappa$.  All clusters with $R_a=0.8 \kpc$ are assumed
depleted and all clusters with energies equal to a circular orbit with
$R_a=20\kpc$ are assumed to be unevolved.  For $R_a=0.8\kpc$, clusters
on circular orbits either evaporate or decay into the nucleus by
dynamical friction (Aguilar, Hut \& Ostriker 1988) and clusters on
eccentric orbits evaporate.  For $R_a=20 \kpc$, the evaporation
timescale on a circular orbit is $70 \Gyr$ for $10^5 \msun$ and is
roughly the same for other orbits of equal energy.  In most cases
observed clusters with $M\leq 10^5 \msun$ initially had masses above
$10^5 \msun$.  In a few cases, extrapolation beyond the initially
defined grid is required.

	In \S\ref{sec:behavior}, we find that low inclination halo
clusters evolve more rapidly than high inclination clusters, although
the differences are not extreme.  However, in order to reduce
computational expense, we neglect low-inclination orbits in the
spherical component and assume that all orbits have high inclination.
This assumption circumvents the roughly factor-of-four increase in
halo phase space grid size required to sample the cylindrical
geometry.  As a result, the initially spherical distribution remains
spherical and we underestimate the amount of evolution about the
midplane of the disk.  The consequences of this assumption are
elaborated below.

\subsection{Galactic model}
\label{sec:MW_model}

	We represent the spherical component of the Galaxy as a
singular isothermal sphere with $V_0=220\kms$ and the disk component
as an exponential disk with radial scale length $R_0=3.5\kpc$
normalized to the disk central density in the solar neighborhood,
$\rho_0=0.15 \msun/\pc^3$ (Bahcall 1984).  The vertical profile is
taken to be Gaussian with a scale height of $z_0=320\pc$.  The
Gaussian was adopted instead of an exponential initially due to
concern about the analyticity of the perturbation and its effect on
adiabatic invariance.  However, comparisons between the two profiles
show no strong differences in heating rate (Weinberg 1994).  The solar
radius is taken to be $R_0=8.5 \kpc$.  All measured radial velocities
are converted to velocities in the Galactic rest frame using an LSR
velocity of $220\kms$ and a solar motion of $(\Pi,\Theta,Z)=(-9,12,7)
\kms$ (Mihalas \& Binney 1981).

\subsection{Parameterization of disk strength}
\label{sec:disk_parms}

	In order to parameterize the strength of the spheroid relative
to the cluster in Paper I, we introduced the quantity $M(x_p)$ which
denotes the fraction of the cluster mass contained within the
pericentric inner Lagrange point.  Here, because of the the disk, it
is convenient to define the following two ratios: $\chi=x_p/R_c$
denotes the ratio of the pericentric inner Lagrange point to the
limiting radius of the cluster; and $\rho_{0,d}(R)/\bar\rho_c$ denotes
the ratio of the disk central density at a given radius $R$ to the
cluster mean density.  This latter parameter defines the tidal
amplitude of the disk relative to the cluster in the same way that
$M(x_p)$ is used to define the relative tidal amplitude of the
spheroid.  Using $\chi$, we can relate the cluster mean density,
$\bar\rho_c$, to the mean density required for tidal limitation in the
spheroid, $\bar\rho_t$, by $\bar\rho_c/\bar\rho_t=\chi^3$.

\section{Results}
\label{sec:results}

\subsection{Physical Behavior}
\label{sec:behavior}

\subsubsection{The importance of disk heating}
\label{sec:disk_heating}

	In Paper I, the evolution of a cluster of fixed mass, $M(x_p)$
and $\kappa$ could be scaled to an orbit of any energy due to the
scale-free nature of the tidal field.  This allowed us to compare the
orbital dependence of cluster evolution in several ways using the same
calculations (c.f. Paper I, \S 3.1).  Adding the disk destroys this
scaling freedom because the quantity $\rho_{0,d}(R_d)/\bar\rho_c$
varies with the radius of disk crossing $R_d$.  In order to compare
cluster evolution on different orbits, we show the mass remaining
after $10 \Gyr$ in tidally limited clusters on orbits of equal
apocenter over a range of mass and eccentricity both with and without
the disk (Tables \ref{tab:disk_mass_rem} and \ref{tab:halo_mass_rem}).

	For equal apocenter and fixed mass, the densities of tidally
limited clusters increase with orbital eccentricity due to the
decrease in $R_t$ with increasing perigalactic angular frequency.
Clusters on eccentric orbits therefore undergo the most rapid
relaxation and have correspondingly large evaporation rates.  As a
result, for evolution in the spheroid alone, the remaining masses of
$10^5 \msun$ clusters decrease monotonically with eccentricity.  Tidal
heating also enhances mass loss rates on low-eccentricity orbits.  The
effect is noticeable in the remaining masses of more slowly relaxing
$10^6 \msun$ clusters.  At high eccentricity, relaxation still
predominates, but at low eccentricity, heating becomes important.  As
a result there is a peak in remaining mass at intermediate
eccentricity.
  
	The large tidal amplitude of the disk relative to the spheroid
greatly enhances heating on low- and intermediate eccentricity orbits
in the disk+sphere calculations.  High-mass low-eccentricity disk
clusters lose equilibrium and disrupt.  The low-mass counterparts do
not disrupt but are rapidly driven to evaporation by strong tidal
stripping.  In addition, since the strength of the disk relative to
the spheroid varies with radius, the importance of disk heating varies
as a function of the radius of disk crossing.  The relative strength
of disk heating is highest at about $8 \kpc$.  As a result, the $10^6
\msun, \kappa=0.9$ disk cluster at $8 \kpc$ loses more mass than does
its counterpart at $4\kpc$ (Table \ref{tab:disk_mass_rem}).  The tidal
effect of the disk diminishes with increasing eccentricity because
cluster densities increase relative to the disk density at the
crossing point.  Disk heating is negligible at $\kappa=0.3$.

\begin{table*}
\caption{Disk clusters: fraction of remaining mass after 10\Gyr}
\label{tab:disk_mass_rem}
\begin{tabular}{lccccl}
\multispan6 \hfil $10^5 \msun$\hfil\\\\
\hline
$\kappa=$&1.0&0.9&0.6&0.3&\\
\hline
4\kpc&0.0&0.0&0.0&0.0&disk+sphere\\
&0.42&0.35&0.0&0.0&sphere\\
8\kpc&0.0&0.0&0.05&0.0&disk+sphere\\
&0.77&0.76&0.40&0.0&sphere\\
\hline\\
\multispan6 \hfil$10^6 \msun$\hfil\\\\
\hline
$\kappa=$&1.0&0.9&0.6&0.3&\\
\hline
4\kpc&0.0&0.54&0.74&0.60&disk+sphere\\
&0.77&0.84&0.89&0.60&sphere\\
8\kpc&0.0&0.25&0.75&0.86&disk+sphere\\
&0.85&0.91&0.95&0.86&sphere\\
\hline
\multispan6 for $5z_0$ oscillation height and $W_0=5$\hfill
\end{tabular}
\end{table*}

	Halo clusters exhibit the same overall tendencies as disk
clusters, but heating rates are lower because resonances are
concentrated at higher frequencies than resonances due to disk
oscillations.  For example, at $90^{\rm o}$ inclination, the $1z_0$
passage time scale is on the order of $1 \Myr$ while the corresponding
period of an oscillation with height greater than $1z_0$ is greater
than $10 \Myr$ for a disk cluster with $R_d>2 \kpc$.  The location of
the resonant frequency match is significant because the increase in
binding energy at higher frequency reduces the response of a stellar
orbit to a perturbation of fixed amplitude.

\begin{table*}
\caption{Halo clusters: fraction of remaining mass after 10\Gyr}
\label{tab:halo_mass_rem}
\begin{tabular}{lccccl}
\multispan6\hfil$10^5 \msun$\hfil\\\\
\hline
$\kappa=$&1.0&0.9&0.6&0.3&\\
\hline
5\kpc&0.14&0.10&0.0&0.0&disk+sphere\\
&0.58&0.55&0.0&0.0&sphere\\
10\kpc&0.74&0.73&0.42&0.0&disk+sphere\\
&0.82&0.82&0.59&0.0&sphere\\
\hline\\
\multispan6 \hfil$10^6 \msun$\hfil\\\\
\hline
$\kappa=$&1.0&0.9&0.6&0.3&\\
\hline
5\kpc&0.49&0.56&0.82&0.72&disk+sphere\\
&0.80&0.86&0.91&0.73&sphere\\
10\kpc&0.80&0.82&0.90&0.90&disk+sphere\\
&0.87&0.93&0.96&0.90&sphere\\
\hline
\multispan6 for $i=90^{\rm o}$ and $W_0=5$\hfill
\end{tabular}
\end{table*}

\subsubsection{Inclination dependence}
\label{sec:inclination}

	The importance of resonance location and frequency matching is
also evident when comparing halo clusters on orbits of different
inclination with respect to the disk.  Table \ref{tab:halo_mass_rem_1}
compares cluster evolution on circular orbits and shows that mass loss
increases as orbits become less inclined because resonances appear at
lower frequency.  At 5\kpc, the best match occurs for $i=30^{\rm o}$,
while at 10\kpc, the best match occurs at $i=15^{\rm o}$.  Radial
differences arise because the disk passage time is fixed at
$1z_0/v_c\sin i$ while the cluster dynamical time scale varies as
$\bar\rho_c^{-1/2}$ as determined by the mean density of the spheroid
through tidal limitation.  The drop in cluster mean density at $10
\kpc$ leads to more efficient heating at lower passage speed.

	The tendency for clusters confined to the disk and at low
inclination to evolve more rapidly implies that an initially spherical
halo distribution will develop a vertically-dependent density profile
with minimum density about the midplane of the disk. This effect is
most pronounced at low eccentricity (\S\ref{sec:disk_heating}) because
the high densities of tidally truncated clusters on highly eccentric
orbits strongly reduce the resonance amplitudes between stellar orbits
and the time-varying disk tidal field.

\begin{table*}
\caption{Halo clusters: fraction of remaining mass after 10 Gyr as a function
		of orbital inclination}
\label{tab:halo_mass_rem_1}
\begin{tabular}{lcccc}
\multispan5\hfil $10^5\msun$\hfil\\\\
\hline
$i=$&$15^o$&$30^o$&$60^o$&$90^o$\\
\hline
5 kpc&0.02&0.0&0.09&0.14\\
10 kpc&0.49&0.59&0.71&0.73\\
\hline\\
\multispan5 \hfil$10^6 \msun$\hfil\\\\
\hline
$i=$&$15^o$&$30^o$&$60^o$&$90^o$\\
5 kpc&0.50&0.41&0.44&0.49\\
10 kpc&0.58&0.69&0.80&0.80\\
\hline
\multispan5 $\kappa=1.0$ and $W_0=5$\hfill
\end{tabular}
\end{table*}

\subsubsection{Density dependence}
\label{sec:density}

\begin{figure}
\epsfxsize=20pc
\epsfbox[12 138 600 726]{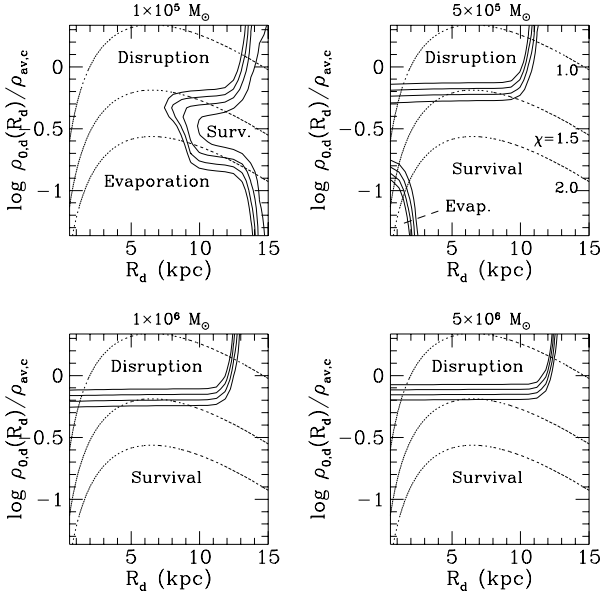}
\caption{Remaining mass at $10 \Gyr$ for disk clusters on circular
	orbits executing $5z_0$ oscillations as a function of the
	ratio of disk central density to cluster central density
	($W_0=5$ profile).  Solid contours show remaining masses in
	the range $3.0\leq \log M_c \leq 4.5$ with $\Delta\log
	M_c=0.5$.  Dotted contours show values of $\chi$ as labeled in
	the top right panel.  Initial masses are given at the top of
	each panel.  Lower density clusters tend to disrupt due to
	tidal heating while higher density clusters tend to evaporate
	due to rapid rates of relaxation.  The density required for
	tidal limitation in the spheroid is too low to allow survival
	against disk heating.}
\label{fig:disk_ev}
\end{figure}

	The rapid disruption of tidally-limited low-eccentricity disk
clusters implies that only those with high initial densities can
survive for long periods after formation.  Figure \ref{fig:disk_ev}
shows survival, disruption and evaporation patterns in clusters on
circular orbits with a range of initial densities, parameterized by
the ratio of disk central density to cluster mean density,
$\rho_{0,d}(R_d)/\bar\rho_c$.  The heat input from the spheroid is
negligible beyond $1 \kpc$ for disk clusters and has therefore been
ignored to provide scaling freedom.  Surviving clusters are bounded by
disruption at low density (small $\chi$)and evaporation at high
density (large $\chi$).  Clusters with $10^5 \msun$ can only survive
for $R_d>7\kpc$.  For higher mass, the evaporation boundary moves to
higher density because of the longer relaxation time scale.  However,
the disruption boundary remains roughly constant at density higher
than the tidal limit.

\subsubsection{Dependence on oscillation height}
\label{sec:z0}

	Table \ref{tab:disk_mass_rem_1} shows the dependence of disk
cluster evolution on oscillation height for clusters on circular
orbits with $\log\rho_{0,d}(R_d)/\bar\rho_c=-0.37$.  This value is
well within the tidal limit set by the spheroid for $R_d>1\kpc$ but
still strongly influenced by disk heating.  Evolution is weakly
dependent on the oscillation amplitude, with the maximum effect
occurring between $2z_0$ and $5z_0$.  We exploit this property below
to remove the vertical dimension from the phase space grid used to
construct the distribution of evolved disk clusters.

\begin{table*}
\caption{Disk clusters: fraction of remaining mass after 10 Gyr as a function
		of disk oscillation height}
\label{tab:disk_mass_rem_1}
\begin{tabular}{lrcccc}
&&\multispan4 \hfil oscillation height\hfil\\
&&$1 z_0$&$2 z_0$&$5 z_0$&$10 z_0$\\
\hline
$10^5 \msun$&4\kpc&0.0&0.0&0.0&0.0\\
	&8\kpc&0.45&0.36&0.38&0.43\\
$10^6 \msun$&4\kpc&0.80&0.76&0.67&0.69\\
	&8\kpc&0.82&0.81&0.77&0.78\\
\hline
\multispan6 $\kappa=1.0$ and $W_0=5$\hfill
\end{tabular}
\end{table*}

\subsubsection{Concentration dependence}
\label{sec:concentration}

\begin{figure}
\epsfxsize=20pc
\epsfbox[12 138 600 726]{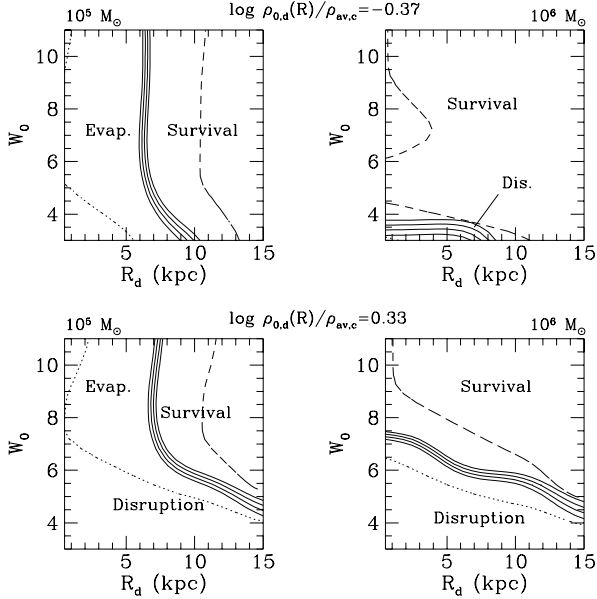}
\caption{Remaining mass after $10 \Gyr$ as a function of initial
	concentration and radius in the disk for clusters on circular
	orbits with indicated masses and mean densities.  Solid
	contours show remaining masses in the range $3.0\leq \log
	M_c\leq 4.5$ with $\Delta\log M_c=0.5$.
	Evaporation/disruption isochrones at $5 \Gyr$ (dotted) and $15
	\Gyr$ (dashed) are also shown.  Clusters with $10^5 \msun$
	evaporate for $R_d\leq 6\kpc$ at high mean density (upper
	left) and evaporate or disrupt for $R_d\leq 7\kpc$ at low mean
	density (lower left).  Clusters with $10^6 \msun$ disrupt for
	$R_d\leq 7\kpc$ and very low concentration at high mean
	density (upper right) and disrupt over a large range in radius
	and concentration for low mean density (lower right).}
\label{fig:conc_ev}
\end{figure}

	The preceding discussion is based on clusters of a single
concentration.  To examine the concentration dependence, we consider
the evolution of disk clusters of varying concentration for a range of
mass and at values of $\rho_{0,d}(R_d)/\bar\rho_c$ which bracket the
range of maximum cluster lifetime.  Figure \ref{fig:conc_ev} indicates
that evaporation dominates at high concentration and high density
while disruption dominates at low concentration and low density.
Clusters show similar trends for increasing eccentricity, with
evaporation becoming more important due to the increasing densities of
tidally limited clusters at fixed Galactocentric radius.

\subsection{Internal properties}

\begin{figure}
\epsfxsize=20pc
\epsfbox[12 138 600 726]{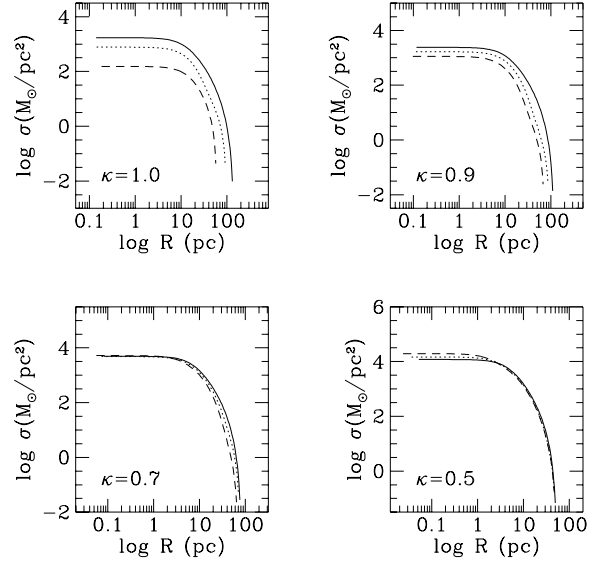}
\caption{The evolution of surface density profiles in four $10^6
	\msun$ clusters on indicated orbits with apogalactica at 8
	\kpc.  Solid lines show initial profiles; dotted lines show
	profiles after $5\Gyr$; dashed lines show profiles after $10
	\Gyr$ except for $\kappa=1.0$ whose profile is shown at
	$7.5\Gyr$, just prior to disruption.  In the strong tidal
	cases ($\kappa=1.0, \kappa=0.9$), halos are truncated and
	profiles develop a steeper fall off than the initial profile.
	In the $\kappa=0.7$ case, no expansion occurs due to the near
	balance between heating and relaxation, but the outer profile
	evolves due to tidal heating.  The $\kappa=0.5$ cluster
	undergoes negligible tidal heating, evidenced by the static
	halo profile, while relaxation leads to increasing central
	densities.}
\label{fig:prof_comp}
\end{figure}

	To illustrate some basic trends in the evolution of internal
cluster properties, we examine traits of the $10^6 \msun$ disk
clusters with apogalactica at $8 \kpc$.  Figure \ref{fig:prof_comp}
compares evolution in projected profiles for cases ranging from tidal
disruption to relaxation-dominated core contraction.  The tidally
dominated $\kappa=1.0$ and $\kappa=0.9$ clusters shows marked
departure from the initial profiles, developing a steeper, roughly
power-law decline.  The limiting radii remain near the expected tidal
boundary which moves inward due to the mass loss.  For $\kappa=0.7$,
heating is strong enough to produce deviation from the initial King
profile, although the central evolution is relatively unaffected.  For
$\kappa=0.5$, heating is so weak that the outer profile remains fixed
while the central regions undergo gravothermal contraction.  The
tidally influenced profiles also show mild concavity in the fall-off,
a feature which resmbles the observed profiles given by Grillmair et
al. (1995,1996), who interpreted their observations as {\it tidal
tails\rm}.  However, the feature evident here is not an unbound tidal
tail but a bound halo region which has been partially cleared through
orbital resonances.

\begin{figure}
\epsfxsize=20pc
\epsfbox[12 138 600 726]{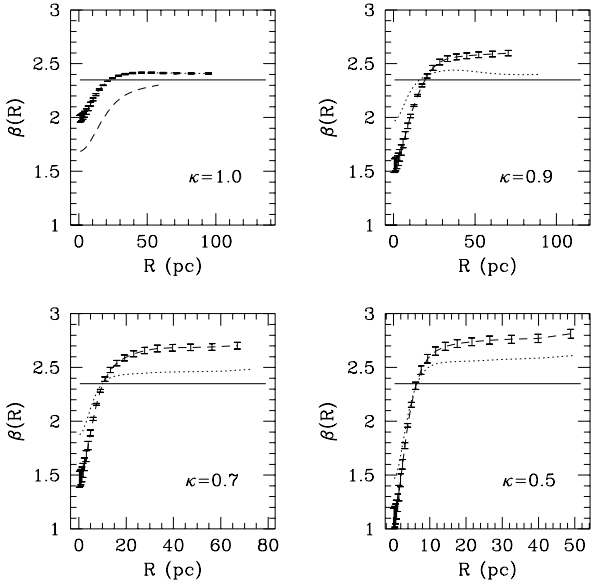}
\caption{The line-of-sight mass spectral index as a function of
	projected cluster radius.  Solid lines show the initial value,
	$\beta=2.35$; dotted lines show the dependence after $5
	\Gyr$; dashed lines show the dependence after $10 \Gyr$,
	except for $\kappa=1.0$ which shows $\beta(R)$ after $7.5
	\Gyr$, just prior to disruption.  Uncertainties are plotted in
	only one case to show the typical size of $1-\sigma$ error
	bars.  The index $\beta(R)$ shows the effect of mass
	segregation in each case.  The increased difference between
	core and halo indices with eccentricity results from the more
	rapid evolutionary timescale at high eccentricity for orbits
	with equal apocenter.}
\label{fig:arp_comp}
\end{figure}

	The evolution of the profile of the mass spectral index is
shown in Figure \ref{fig:arp_comp}.  Mass segregation occurs in every
case, regardless of the strength of tidal heating.  Most low-mass
stars evaporate while the $\kappa=1.0$ cluster disrupts as indicated
by the strong flattening of $\beta(R)$ at all radii.  The other cases
show flattening of the spectrum in the core and steepening in the halo
with differences that increase with eccentricity.  The increasing
differences result from the shorter evolutionary time scales at high
eccentricity for orbits with equal apocenter.  Aside from differences
in time scale, the evolution of the mass spectral index does not
depend significantly on orbit.  The spectral index remains
approximately constant in time near the initial half-mass radius of
the cluster.  However, the half-mass radius is relatively constant
only in the most eccentric cases and undergoes considerable evolution
whre tidal effects are strong.

\subsection{Characteristics of the present-day cluster population}
\label{sec:pdchar}

\subsubsection{Distribution of cluster masses}
\label{sec:mass_spect}

	The mass spectrum of observed clusters is not well-described
by a single power-law index over the mass range considered here.
Harris \& Pudritz (1994) find a change in slope near $10^5 \msun$.
This trend is evident in Table \ref{tab:mass_comp} which compares fits
to the mass spectrum using three different models: a single power law,
a two-component power law and a Gaussian magnitude distribution.  The
single power law shows a relatively flat spectrum in agreement with
Harris \& Pudritz (1994).  The two-component power law shows a fairly
steep dependence for $M>2.2\times 10^5 \msun$ and a nearly flat
spectrum for masses below that.  The Gaussian magnitude distribution
peaks at $2.7\times 10^5 \msun$, consistent with the two-component
power law.

	Likelihood ratio tests show that the single power law can be
rejected in favor of both the Gaussian magnitude distribution and
two-component model at better than 99\% confidence.  The Gaussian and
two-component power law can be discriminated with only 40\%
confidence.  However, because the data is so sparse, we adopt the
single power law as the simplest model which provides a tenable
description of the overall cluster mass distribution.

\begin{table*}
\caption{Models of present-day mass spectrum}
\label{tab:mass_comp}
\begin{tabular}{lccccccc}
Model&$\alpha_1$&$\sigma_{\alpha_1}$&$\alpha_2$&$\sigma_{\alpha_2}$
	&$M_{cut} (\msun)$&$\sigma_{M_{cut}}$&log L\\
\hline
power-law&0.80&0.05&-&-&-&-&-260.56\\
two-component\\
\hfil power-law&0.034&0.15&1.62&0.15&$2.2\times10^5$&$2.6\times10^4$&-248.38\\
\hline\\\\
\end{tabular}
\begin{tabular}{lccccc}
Model&$V_0$&$\sigma_{V_0}$&$\sigma_V$&$\sigma_{\sigma_V}$&log L\\
\hline
gaussian mag.&-7.56&0.11&1.30&0.10&-251.03\\
\hline
\end{tabular}
\end{table*}

\subsubsection{Orbital dynamics and spatial distribution}
\label{sec:orb_dyn}

\begin{figure}
\epsfxsize=20pc
\epsfbox[12 138 600 726]{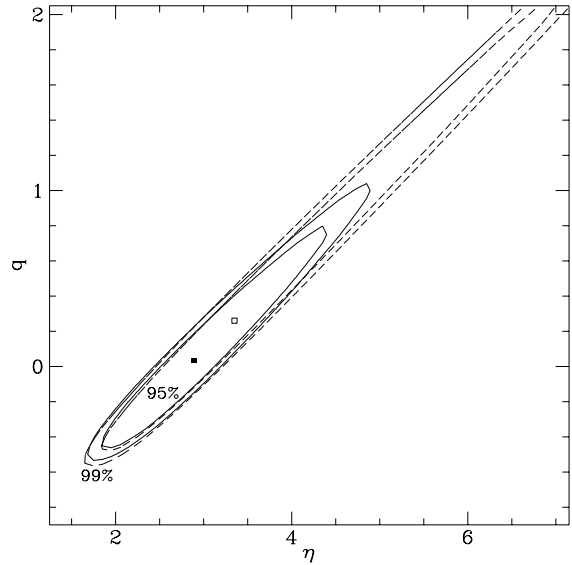}
\caption{Comparison of Mestel sphere fits to data with (solid) and 
	without (dashed) available three-space velocities.  The
	additional information in the augmented data set provides some
	constraint on the degree of tangential anisotropy compared to
	the pure radial velocity data.  The uncertainty still
	dominates, but we can rule out strongly anisotropic
	distributions.}
\label{fig:data_comp}
\end{figure}

	Figure \ref{fig:data_comp} shows the inferred model parameters
for the Mestel sphere both with and without the tangential velocities.
The comparison shows the improved constraints that the additional
information provides.  With the radial velocity data set, the
tangential anisotropy of the Mestel sphere fit is unconstrained; in
the augmented data set, the confidence contours close within $q=1.0$
and $\eta=5.0$ at 99\% confidence.  The contours are centered about
isotropy and rule out strong anisotropy.  We will use the augmented
data set throughout the remainder of this paper.

	The estimated value of the anisotropy radius of the Eddington
sphere, $R_a=20 \kpc$ (Table \ref{tab:fit_comp}), also indicates an
isotropic distribution.  A likelihood ratio test weakly favors the
Eddington sphere over the Mestel sphere with 75\% confidence.  The
better fit results primarily because the Eddington sphere has a
central core and a steep decline for $R>20\kpc$.  However, as shown in
Paper II, cluster distributions probably follow steep power-law
profiles initially and develop cores at later times through dynamical
evolution.  Both models can provide nearly coreless power-law density
profiles, but the Eddington sphere becomes extremely radially biased
in this case.  The Mestel sphere, by contrast, can have arbitrary
orbital anisotropy for a given spatial profile.  Therefore,
considering the relative isotropy of the present-day population, the
Mestel sphere provides a more realistic description of an initial
cluster population.  To investigate initial conditions below, we use
only the Mestel sphere to model the spherical portion of the cluster
distribution.

\begin{table*}
\caption{Comparison of fits to Mestel and Eddington spheres.}
\label{tab:fit_comp}
\begin{tabular}{lccccccc}
Model&$\eta$&$\sigma_{\eta}$&$q$&$\sigma_q$&$r_a$&$\sigma_{r_a}$&log L\\
\hline
Mestel&2.89&0.25&0.04&0.11&-&-&-3544.4\\
Eddington&2.54&0.10&-&-&20.1&4.1&-3542.3\\
\hline
\end{tabular}
\end{table*}

	The two-component analysis implies that 96\% of the clusters
are in a spherical component, according to the estimated value of $F$
(Table \ref{tab:an_sd}).  The inferred sphericity of the distribution
is in agreement with previous analyses of the full cluster system
(Frenk \& White 1980; Thomas 1989; Chernoff \& Djorgovski 1989).
However, a likelihood ratio test rejects both the Mestel and Eddington
spheres in favor of the two-component model at better than 99\%
confidence, indicating that the purely spherical models inadequately
represent the dynamics of the full system.  The preference stems from
the strong tangential anisotropy attributed to the disk component.

	Strong tangential anisotropy is expected in a disk component
due to the presence of the rapidly rotating system of high-metallicity
disk clusters (Armandroff 1989).  However, our estimate implies that
only 4 clusters in the sample are associated with the disk, while
recent determinations indicate the presence of nearly 25 metal-rich
disk clusters (Armandroff 1993).  This apparent contradiction may
result from the obscuration of low-latitude clusters and the greater
evolutionary rate of clusters near the disk (\S\ref{sec:inclination}),
both of which lead to a deficit of halo clusters near the disk.  As a
result, the `hole' in the spherical component is filled by `borrowing'
the isotropic portion of the disk system which is kinematically
well-matched to the halo system.  This leaves a residual system of
disk clusters with very high angular momentum.

\begin{table*}
\caption{Present-day two-component model}
\label{tab:an_sd}
\begin{tabular}{lccccccccc}
&$\eta$&$q$&$\alpha$&$\eta_d$&$q_d$&$\eta_z$&$\alpha_d$&$F$& log L\\
\hline
estimate&2.75&-0.035&0.76&17.0&7.8&0.37&1.04&0.96&-3495.00\\
uncertainty&0.24&0.12&0.05&7.08&3.34&0.16&0.13&$^{+0.04}_{-0.10}$&-\\
\hline
\end{tabular}
\end{table*}

\subsection{Initial conditions of the cluster population}
\label{sec:init}

\begin{figure}
\epsfxsize=20pc
\epsfbox[12 138 600 726]{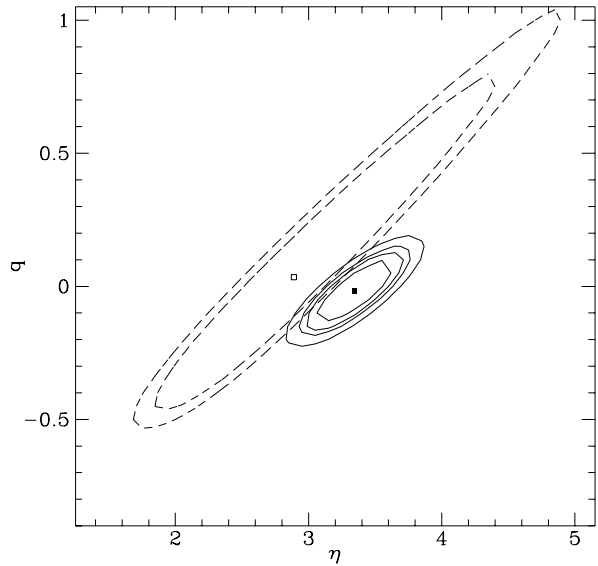}
\caption{Comparison of 50\%,90\%,95\%,99\% confidence levels in
		estimated initial parameters (solid) with levels in
		estimated present-day parameters superimposed from
		previous figure (dashed).  Constraints are stronger on
		the initial distribution and indicate initial
		isotropy.  The slope of the initial density
		distribution $-(\eta+2q)$ is approximately $r^{-3.35}$
		while the present-day slope is best described as
		$r^{-2.95}$.}
\label{fig:ev_comp}
\end{figure}

	Evolution of the pure Mestel sphere produces a profile which
is shallower at present than in the past (Figure \ref{fig:ev_comp})
due to the more rapid evolution at small galactocentric radii.  The
initial velocity distribution is tightly constrained to isotropic.

	Table \ref{tab:ev_sd_comp} compares the best-fit parameters of
the spherical model with the two-component model.  The spheroid in the
two-component model has a slightly shallower decline but is
considerably more radially biased because higher angular momentum
orbits are mainly associated with the disk.  The uncertainties in the
disk parameters are reduced because the present-day population
fraction (1-$F$) is larger than in the present-day model.  The current
population is estimated to number about 16 clusters. This is still
somewhat below the expected size, but again may be the result of
imposing spherical symmetry on the evolved halo distribution.  A
likelihood ratio test rejects the purely spherical model with 97.5\%
confidence.

	The analysis predicts that the velocity distributions of both
halo and disk components become more tangentially biased with time due
to the more rapid evaporation of clusters on eccentric orbits.  The
initial orbit distribution of the halo component has fairly strong
radial bias, with approximately 41\% of its kinetic energy in radial
motion, while the disk component has a nearly isotropic initial orbit
distribution.  In addition, the initial halo cluster density
distribution has power law index of $\eta-2q=3.38$, while the disk
cluster density distribution has power law index $\eta_d-2q_d=2.25$.

\begin{table*}
\caption{Comparison of initial conditions in 2-component and spherical models}
\label{tab:ev_sd_comp}
\begin{tabular}{lcccccccc}
&\multispan8\hfil spherical \hfil\\
&$\eta$&$q$&$\alpha$&$\eta_d$&$q_d$&$\alpha_d$&$F$& log L\\
\hline
estimate&3.35&-0.018&0.95&-&-&-&-&-3486.94\\
uncertainty&0.32&0.16&0.06&-&-&-&-&-\\
\hline\\
&\multispan8\hfil 2-component \hfil\\
&$\eta$&$q$&$\alpha$&$\eta_d$&$q_d$&$\alpha_d$&$F$& log L\\
\hline
estimate&2.82&-0.28&0.96&2.38&0.066&1.15&0.89&-3478.79\\
uncertainty&0.27&0.03&0.07&0.53&0.10&0.23&0.08&-\\
\hline
\end{tabular}
\end{table*}

\begin{figure}
\epsfxsize=20pc
\epsfbox[12 138 600 726]{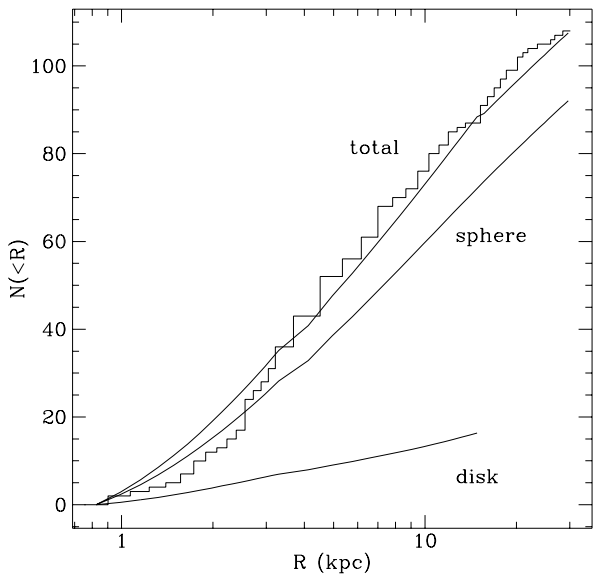}
\caption{Comparison of the evolved two-component model with the
	observed cumulative distribution of clusters with $R<30\kpc$
	in the present sample.  The total number of clusters is 108.
	The spherical component is estimated to have 92 clusters and
	the disk 16.}
\label{fig:ev_sd}
\end{figure}

	Figure \ref{fig:ev_sd} compares the cumulative distribution of
clusters in our data sample with the evolved profile of the
two-component model along with the separate contributions of the disk
and sphere.  The model matches the data fairly well.  At small radii
our models overestimate the expected number of clusters.  However, the
KS confidence that the observed and model distribution differ is only
58\%.

\begin{figure}
\epsfxsize=20pc
\epsfbox[12 138 600 726]{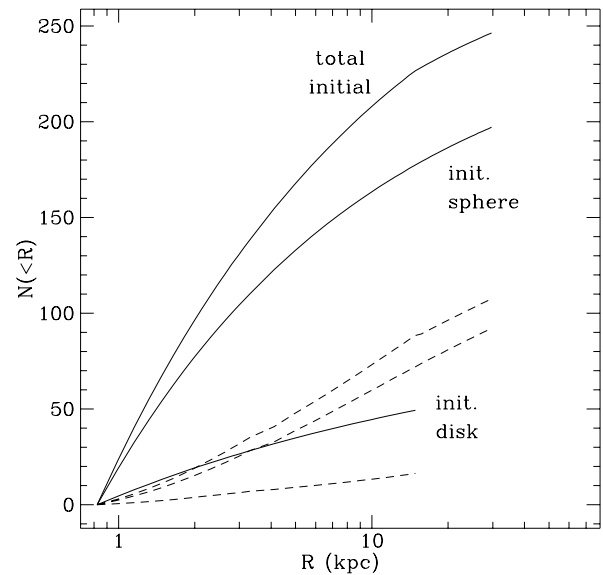}
\caption{Comparison of initial (solid) and evolved (dashed- from
	previous figure) two-component models.  The total initial
	number of clusters is estimated to be about 250 with 200 in
	the spheroid and 50 in the disk.}
\label{fig:ev_sd.1}
\end{figure}

	Using the initial conditions from the two-component model, we
derive the estimated initial distribution of clusters (Figure
\ref{fig:ev_sd.1}).  The total initial population has roughly 250
clusters, with 200 in the spheroid and 50 in the disk.  Approximately
43\% of the initial population remains.

\section{Discussion}
\label{sec:discussion}

	Before significant mass loss through stellar evolution, young
clusters typically have strongly concentrated profiles (e.g. Elson,
Fall \& Freeman 1987).  During this phase, concentration is reduced as
mass loss dominates gravothermal contraction, driving the expansion of
individual clusters (Chernoff \& Weinberg 1990).  When stellar
evolution mass loss subsides---and our model calculations begin---the
resulting population will consist of clusters with a range of
profiles.  The calculations shown in \S\ref{sec:concentration}
indicate, however, that evaporation times do not strongly depend on
initial concentration.  High concentration clusters undergo weaker
tidal heating but maintain correspondingly larger rates of relaxation,
so that evaporation times, for given mass, are approximately
independent of concentration.  The weak dependence of evaporation time
on density is similar (\S\ref{sec:density}).  The results of our
calculations, therefore, do not depend significantly on the
distribution of initial concentration.

	Estimates of initial population size are derived assuming that
an initially spherical cluster distribution remains spherical.
However, clusters on orbits confined near the disk evolve more rapidly
than those on high inclination orbits, leaving an axisymmetric
distribution with minimum density about the midplane of the disk.
Estimates of initial halo cluster population size should not change
significantly when accounting for additional cluster loss at low
inclination because only $\sim 10\%$ more depletion of halo clusters
is required to account for the remaining disk clusters.  However, the
observed disk cluster population is about 60\% larger than our
estimate of 16.  Assuming that our estimate for the initial population
size scales accordingly, we estimate that the disk population may have
had roughly 80 members initially.

	The choice of the mass spectrum of clusters also influences
estimates of initial population size.  While use of a single power law
is necessitated by sample size, the two-component power law clearly
provides the best fit.  To gauge the importance of this choice, we use
the two-component fit and the results of Paper II as a guide.  We
estimate that the spectral index may decrease by $0.2$ for $M\gta
M_{cut}$ and by $0.3$ for $M\lta M_{cut}$ since the high mass range is
very similar to that considered in Paper II while more rapid evolution
occurs in the low mass range.  This implies initial spectral indices
of $\alpha_1=0.3$ and $\alpha_2=1.8$ and 49\% of the population with
$M\leq M_{cut}$.  In the single index fit with $\alpha=0.95$, 47\% of
the population has $M\leq M_{cut}$.  Since most of these clusters
evaporate within a Hubble time and since they dominate the depleted
population, estimates of initial size will not change substantially
when using a two-index model.

	Both the spatial and kinematic distributions of halo globular
clusters differ from those of halo field stars.  Harris (1976) and
Zinn (1985) found that the spatial distribution of the full cluster
system goes as $R^{-3.5}$ for $4\leq R\leq20 \kpc$, in good agreement
with the profile of halo field stars (Freeman 1996).  At smaller radii
the profile becomes shallower and flattens into a core (Ostriker,
Binney \& Saha 1989).  Similarly, while the distribution of halo field
star orbits appears to have significant radial bias (Beers \&
Sommer-Larsen 1995), the distribution of halo cluster orbits appears
to be nearly isotropic (\S\ref{sec:orb_dyn}).  Our results from the
present-day fits also exhibit the discrepancy between the spatial
distributions of halo clusters and halo field stars.  The power-law
profile derived using the Mestel sphere appears considerably flatter
than $R^{-3.5}$ because we have included clusters with $R\leq 4\kpc$
in the data set.  The Eddington sphere best represents the
distribution over the full radial range.  It has a core, density
$\rho\propto R^{-2.5}$ at small radius, a steep drop beyond $20 \kpc$
with density $\rho\propto R^{-4.5}$ and the expected power-law decline
with density $\rho\propto R^{-3.5}$ at intermediate radii.

	Based on the estimates of initial conditions, we conclude that
cluster evolution can account for the difference between the spatial
distributions of halo clusters and field stars.  The estimated initial
density profile $\rho\propto R^{-3.38\pm0.3}$ in both spherical and
two-component models over the full radial range of the data.  This is
in good agreement with the halo field star profile $\rho\propto
R^{-3.29\pm0.24}$ recently determined by Sommer-Larsen \& Zhen (1990)
as well as the conventionally accepted value of $R^{-3.5}$.

	Evolution also accounts for at least some differences in halo
cluster and field star kinematics.  As discussed in \S\ref{sec:init},
the predicted initial halo cluster population in the 2-component model
has a strong radial bias which diminishes over time due to the
preferential evaporation of clusters on eccentric orbits.
Approximately $41\%$ of the kinetic energy of this component is
initially in radial motions.  Compared to $33\%$ implied by the
isotropy of the observed halo cluster distribution, this is closer to
the value of $54\%$ derived for halo field stars by Beers \&
Sommer-Larsen (1995) and is roughly consistent with the value of
$44\%$ derived by Norris (1986).

	The predicted initial disk population, by contrast, is unlike
the observed stellar thick disk.  The estimates indicate that the
present-day high angular momentum population developed from a nearly
isotropic initial distribution through the selective evaporation of
clusters on eccentric orbits.  Although the model underestimates the
size of the disk population, any resulting kinematic bias predicts
higher angular momentum in the system because lower angular momentum
members cannot be distinguished from the halo clusters.  Evidence for
a flattened component with nearly random kinematics may exist in the
sample of halo clusters studied by Zinn (1993) and in the halo field
star sample studied by Sommer-Larsen \& Zhen (1990).

	Overall, the results of this paper point to a scenario which
correlates the origin of halo field stars and clusters during Galaxy
formation.  The predicted disk cluster system is less clearly
correlated with observed disk stellar populations but may be related
to an intermediate phase of the dissipative collapse which is thought
to have given rise to the disk (Larson 1990; Zinn 1993).  The results
do not contradict merger-induced heating of an initially cold disk
(Quinn, Hernquist \& Fullager 1993), provided that initially circular
disk cluster orbits can be isotropized by the accretion event.

	Finally, we emphasize that our predictions describe the
initial population of clusters which evolve quasi-statically through
relaxation and tidal heating.  Since clusters probably formed with a
range of initial densities, some would have been prey to rapid tidal
disruption (Paper I).  However, the significance of such initial
conditions cannot be divined from the observed population because,
excepting a few clusters such as Pal 5 (Cudworth 1993), the majority
of observed clusters must have formed with initial densities which
allowed quasi-static evolution and long-term survival.  Consequently,
the importance of tidal disruption in shaping the cluster population
at an early epoch can be described only through models of the initial
conditions of the cluster system in a proto-Galactic or cosmological
context (Paper I).

\section{Conclusions}
\label{sec:conclusions}

	The main conclusions of this work are as follows:

\begin{enumerate}

	\item The disk dominates tidal heating on low-eccentricity orbits.
	The effect of the spheroid is negligible except for $R\lta 1\kpc$.
		
	\item Disk oscillations heat more efficiently than disk
	shocking due to better frequency matching in typical cluster
	potentials.

	\item Cluster evolution depends weakly on initial
	concentration and density due to the combined effects of tidal
	heating and evaporation

	\item The evolution of disk clusters depends weakly on
	oscillation height.

	\item The loss of clusters on low-inclination orbits implies
	that the distribution of halo clusters is less dense near the
	disk.

	\item The evaporation of clusters on high-eccentricity orbits
	leads to increasing tangential bias in the distribution of
	cluster orbits.

	\item The estimated initial spatial distribution of halo
	clusters matches the present-day distribution of halo field
	stars.  The estimated initial kinematic distribution is nearer
	to the observed radial anisotropy in the kinematics of halo
	field stars, having approximately 40\% of its energy in radial
	motions.

	\item The estimated initial distribution of disk clusters does
	not match the kinematic distribution of the stellar disk.
	However, it is similar to the flattened halo samples studied
	by Zinn (1993) and Sommer-Larsen \& Zhen (1990).

	\item The outer profiles of evolving clusters deviate from the
	initial profiles in cases of moderate to strong tidal heating,
	leading to considerable evolution in the half-mass radius.
	Profiles may also show density inflections which result from
	resonant clearing and which resemble the observed profiles of
	Grillmair et al. (1995,1996).

	\item Profiles of the mass spectral index become flatter in
	the core and steeper in the halo due to mass segregation.  The
	profiles evolve similarly on all orbits and differ at fixed
	times only through orbitally determined differences in
	evolutionary timescale.

\end{enumerate}

\section*{Acknowledgements}
This work was supported in part by NASA award NAGW-2224.

\appendix

\section{Heating rate for disk oscillations}
\label{sec:disk}

	Clusters confined to the disk oscillate about the midplane and
are heated through resonance with the periodic tidal compression in
the direction perpendicular to the disk plane.  As discussed in
Weinberg (1994), the tidal potential of the disk is well-approximated
by the leading term of an expansion of the disk potential about the
center of mass of the cluster:

\begin{equation}
H_1=2\pi G\rho\bigl[Z(t)\bigr]z^2, 
\end{equation}

\noindent where we have substituted the density for the second
derivative of the disk potential using Poisson's equation.  The
quantity Z(t) is the cluster position as a function of time, and z
refers to the position of a star relative to the center of the
cluster.
 
        Expanding the $z^2$ factor in the action-angle Fourier series as
defined in Tremaine \& Weinberg (1984), we obtain
 
\begin{equation}
z^2=\AAsum\bigl[{2\over 3}{\sqrt {4\pi\over 5}}V_{2 l_2 0}(\beta)
        +{1\over 3}\sqrt{4\pi}V_{0 l_2 0}(\beta)\bigr ]
        X^{l_1}_{l_2}e^{i\ldw}.
\end{equation}
 
\noindent Substituting the resulting expression for $H_1$ into
equation (5) of Paper I, we derive the heating rate

\begin{eqnarray}
\Ebar=-8\pi^4P{df\over dE}\AAsum\delta_{l_3 0}\bigl( {1\over15}+
        {1\over5}\delta{l_2 0}\bigr)(\ldomega)^2|X^{l_1}_{l_2}|^2 \nonumber\\
        \Fsum |a_n|^2 \delta(\ldomega-n\omega).
\end{eqnarray}
 
\noindent Choosing the vertical profile of the disk and the amplitude
of vertical oscillations completely specifies the rate of energy input
for given cluster profile.  In the present work, we adopt a Gaussian
vertical profile for the disk.  Comparison with an exponential
vertical profile reveals little difference in overall heating rate at
any oscillation amplitude.

\section{Coordinate systems}
\label{sec:coordinates}

	Our analysis uses three coordinate systems: heliocentric
spherical coordinates, Galactocentric spherical coordinates and
Galactocentric cylindrical coordinates.  The first system is the
natural observational reference frame while the second and third are
the natural reference frames for the sphere and disk models,
respectively.  In the heliocentric frame, we use $(r,l,b)$ or
distance, Galactic longitude and Galactic latitude.  In the
Galactocentric spherical frame, we use the $(R,\Phi,\Theta)$ to denote
Galactocentric distance, colatitude and azimuth respectively.  In the
Galactocentric cylindrical frame, we use the $(R_d,\Phi,Z)$ to denote
Galactocentric radius in the disk, azimuth and height above the plane,
respectively.

	Velocity components in a particular direction are denoted with
the corresponding subscript.  In the heliocentric frame,
$(v_r,v_l,v_b)$ are the radial, azimuthal and latitudinal components,
respectively.  In the Galactocentric spherical frame,
$(v_R,v_{\Phi},v_{\Theta})$ are the radial, azimuthal and latitudinal
components, respectively.  In the Galactocentric cylindrical frame,
$(v_{R_d},v_{\Phi},v_Z)$ are the polar radial, azimuthal and vertical
velocities, respectively.  We also write the components in the
equivalent inner product form so that, for example, the Heliocentric
velocity components are $(\vec v\cdot\widehat r,\vec v\cdot\widehat
l,\vec v\cdot\widehat b)$.

\section{Generalized Mestel Disk}
\label{sec:mestel_disk}

	The phase space distribution function for disk clusters in the
thin disk approximation has the form

\begin{equation}
{\partial N\over \partial E_d\partial J_z^2\partial E_z}=f_d(E_d,J_z^2)g(E_z)
\end{equation}

\noindent where $f_d(E_d,J_z^2)$ governs the distribution in the plane
of the disk, $g(E_z)$ governs the distribution perpendicular to the
disk, $E_d$ denotes orbital energy in the disk, $J_z$ refers to the
angular momentum along the Z-axis and $E_z$ denotes the energy of
vertical osillations.

	The Mestel disk distribution is defined as

\begin{equation}
f_d(E_d,J_z^2)=Ae^{-E_d/\sigma_d^2}J_z^{2q_d},
\end{equation}

\noindent where $\sigma_d^2$ is the isothermal velocity dispersion in
the disk.  The radial and azimuthal velocity dispersions are
$\sigma_{R_d}^2=\sigma_d^2$ and $\sigma_{\Phi}^2=(2q_d+1)\sigma_d^2$
which implies $q_d=\sigma_{\Phi}^2/2\sigma_{R_d}^2-{1\over 2}$, so
that $-{1\over 2}\leq q_d \leq \infty$.  The quantity $q_d$ defines
the anisotropy of the orbit distribution in the disk.  For
$q_d=-{1\over 2}$, the distribution has only radial orbits while as
$q_d\to\infty$ the distribution has only circular orbits (Henon 1973;
Barnes et al. 1986).

	The isothermal vertical distribution is defined as

\begin{equation}
g(E_z)=Be^{-E_z/\sigma_z^2}
\end{equation}

\noindent where $\sigma_z^2$ is the isothermal vertical velocity
dispersion.  In the Milky Way, $\sigma_z^2$ varies with radius in the
disk because the scale height is constant while the midplane density
varies.  Assuming it to be fixed introduces some bias into the
expected vertical velocities at a given radius.  However, this has no
effect on our conclusions because we find that heating by disk
oscillations is nearly independent of oscillation amplitude (or,
equivalently, velocity at the midplane; c.f. \S\ref{sec:behavior}).

	Integrating over $v_{R_d}$, $v_{\Phi}$, $v_Z$ and $Z$ yields the
surface density in the logarithmic potential

\begin{equation}
{dN\over d^2R}=C{\sqrt\pi}(2\sigma_d)^{q+1}\Gamma(q+{1\over 2})
	R_d^{-(\eta_d-2q_d)}
\end{equation}

\noindent where we absorb all vertical integration constants into the 
factor C and define the parameter $\eta_d=v_c^2/\sigma_d^2$.  For
$\eta_d-2q_d=2$, the density goes as $\ln R_d$.

\section{Generalized Mestel Sphere}
\label{sec:mestel_sphere}

	The phase space distribution function for halo clusters has
the form

\begin{equation}
{\partial N\over \partial E\partial J^2}=f(E,J^2)
\end{equation}

\noindent where $E$ is the total energy and $J$ is the total angular 
momentum of a cluster.

	The Mestel sphere has the same form as the Mestel disk,

\begin{equation}
f(E,J^2)=Ae^{-E/\sigma^2}J^{2q},
\end{equation}

\noindent but is a three-dimensional distribution so that the radial
velocity dispersion $\sigma_R^2=\sigma^2$ and tangential velocity
dispersion $\sigma_T^2=2(q+1)\sigma^2$.  Consequently,
$q=\sigma_T^2/2\sigma_R^2-1$ and $-1\leq q\leq\infty$. The quantity
$q$ defines the anisotropy of the orbit distribution.  For $q=-1$, the
distribution has only of radial orbits while as $q\to\infty$ the
distribution has only circular orbits (Henon 1973; Barnes et
al. 1986).  Integrating over $v_R$, $v_{\Phi}$ and $v_{\Theta}$ gives
the volume density in the logarithmic potential

\begin{equation}
{dN\over d^3R}=A\pi^{3/2}(2\sigma^2)^{q+3/2}\Gamma(q+1)R^{-(\eta-2q)}
\end{equation}

\noindent where we define $\eta=v_c^2/\sigma^2$.  For $\eta-2q=3$, 
the density goes as $\ln R$.

\section{Generalized Eddington sphere}
\label{sec:eddington_sphere}

	The Eddington model has the distribution function

\begin{equation}
f(E,J^2)=Ae^{-E/\sigma^2}e^{-J^2/2R_a^2\sigma^2},
\end{equation}

\noindent which implies that $\sigma_R^2=\sigma^2$ and 
$\sigma_T^2=2\sigma^2/(1+R^2/R_a^2)$.  Integrating over velocities
gives the volume density in the logarithmic potential

\begin{equation}
{dN\over d^3R}=A2\pi(2\pi\sigma^2)^{3/2}{R^{-\eta}\over{1+R^2/R_a^2}}
\end{equation}

\noindent where we define $\eta=v_c^2/\sigma^2$.

\section{Likelihood with incomplete data sets}
\label{sec:likelihood}

	An incomplete data set is defined as one in which some
observations are missing.  For each cluster, the observations have
varying completeness with respect to the full set of observations
required in a given model.  There are several approaches to deriving
parameter estimates in this situation.  In the present work, we adopt
a likelihood-based estimation scheme\footnote{The following discussion
is based on the presentation of Little \& Rubin (1987), chapter 5.}.

	If we denote the full phase-space vector for our models $(\vec
r,\vec v,m)$ as $Y$ and write $Y=(Y_{obs},Y_{mis})$, where $Y_{obs}$
signifies the observed data and $Y_{mis}$ signifies the missing data,
then $f(Y\vert\vec\theta)=f(Y_{obs},Y_{mis}\vert\vec\theta)$ denotes
the underlying probability of observing all quantities $Y_{obs}$ and
$Y_{mis}$, where $f(\cdot\vert\vec\theta)$ is governed by the
parameters $\vec\theta$.  Integrating the distribution over each case
of missing data gives the marginal probability density of $Y_{obs}$:

\begin{equation}
\label{eq:marginal_d}
f(Y_{obs}\vert\theta)=\int f(y_{obs},Y_{mis}\vert\vec\theta)dY_{mis}.
\end{equation}

\noindent For independent observations, we denote the marginal probability
for the i$^{th}$ cluster as $f_i=f(Y_{i,obs}\vert\vec\theta)$ and write
the likelihood function in the usual way as the joint probability of
the observations given the model:

\begin{equation}
\label{eq:lhood}
L(\vec\theta)=\prod_i f_i.
\end{equation}

	Using $L(\vec\theta)$ to derive inferences concerning
$\vec\theta$ requires that there are no selection effects leading to
the systematic absence of data for a particular class of observations.
In practice, of course, we know that this is not the case for
observations of globular clusters, where, for example,
latitude-dependent extinction results in the absence of radial
velocities.  However, in the present analysis we make no attempt to
derive any type of selection function.  One approach which does not
ignore selection effects is the {\it Expectation-Maximization\rm}
algorithm, an iterative procedure which provides estimates for the
model parameters as well as the missing data (Little \& Rubin 1987,
ch. 7).

	The standard data set used in analyses of the spatial
distribution and kinematics of the cluster system consists of cluster
positions, masses and heliocentric radial velocities (Aguilar, Hut \&
Ostriker 1988; Thomas 1989).  For example, if we take a spherical
model for the cluster distribution function $f(E,J^2)$, the marginal
probability of any given observation is

\begin{eqnarray}
\bar f(v_r,R)=\int dv_ldv_bf({1\over 2}(v_r^2+v_l^2+v_b^2)+\Psi(R),\nonumber\\
	R^2((\vec v\cdot\widehat\Phi)^2+(\vec v\cdot\widehat\Theta)^2)), 
\end{eqnarray}

\noindent where $\Psi(R)$ is the potential, and the tangential velocity
components are written in inner product notation.

\end{document}